\documentclass[preprint,proceedings]{rmaa}

\newcommand{\xmm}{{\it XMM-Newton}}
\newcommand{\epic}{{\it EPIC}}
\newcommand{\pn}{{\it pn}}
\usepackage{amsfonts}
\SetYear{2007}
\SetConfTitle{The Nuclear Region, Host Galaxy and Environment
 of Active Galaxies}

\title{X-ray properties of a sample of polar-scattered Seyfert galaxies}

\author{  E. Jim\'enez-Bail\'on\altaffilmark{1,2}, M. Guainazzi\altaffilmark{3}, G. Matt\altaffilmark{4}, S. Bianchi\altaffilmark{4}, Y. Krongold\altaffilmark{1}, E. Piconcelli\altaffilmark{5}, M. Santos LLe\'o\altaffilmark{3} \and N. Schartel\altaffilmark{3}}

\altaffiltext{1}{Instituto de Astronom\'\i{}a, Universidad Nacional
  Aut\'onoma de M\'exico, Apartado Postal 70-264, 04510-Mexico DF, M\'exico }
\altaffiltext{2}{LAEFF-INTA, Apdo 50727, 28080-Madrid, Spain.}
\altaffiltext{3}{XMM-Newton Science Operation Centre, ESAC, ESA, Aptdo. 50727, 28080-Madrid, Spain}
\altaffiltext{4}{Universita Roma Tre, Via della Vasca Navale 65, I-00146, Roma, Italy}
\altaffiltext{5}{Osservatorio Astronomico di Roma, Via Frascati 33, I-00040, Monteporzio Catone, Italy}

\suppressfulladdresses

\listofauthors{E. Jim\'enez-Bail\'on, M. Guainazzi, G. Matt, S. Bianchi, Y. Krongold, E. Piconcelli}
\indexauthor{Jim\'enez-Bail\'on, E.; Guainazzi, M.;  Matt,G.;  Bianchi, S.; Krongold, Y.; Piconcelli, E.}

\addkeyword{Galaxies:Active}
\addkeyword{Galaxies: Nuclei}
\addkeyword{Accretion Disks}
\addkeyword{X-ray: Galaxies}

\begin{document}
% Typeset article header
\maketitle 

\boldabstract{  We present  the  results on  an XMM-Newton  systematic
  analysis of a  sample of nine Seyfert 1  galaxies.  When observed in
  polarised light, the spectra of  the selected sources are similar to
  those of Seyfert 2 galaxies. This peculiarity strongly suggests that
  these AGN are  viewed with an inclination comparable  with the torus
  opening angle.   Our results are consistent with  this scenario and,
  taking advantage  of this favourable geometrical  condition, we were
  able  to  investigate in  detail  the  physical  properties and  the
  distribution of the circumnuclear gas in these sources.}

\section{Introduction}

The detection of broad polarised lines in Seyfert~2 nuclei stands that
this type  of AGNs are  intrinsically the same objects  than Seyfert~1
galaxies and the differences  in their observational properties can be
 understood as orientation  effects.  Within the Unified Models,
the central continuum  source of AGNs and the  broad line region (BLR)
are surrounded  by an optically  and geometrically thick  structure of
dust  and molecular gas,  likely following  a toroidal  geometry.  The
orientation with respect to our line of sight of Seyfert~2 galaxies is
such  that  emission  from  the  central engine  is  shielded  by  the
molecular torus  and therefore its continuum and  broad emission lines
are only observed in polarised light as a result of scattering outside
the torus.

Firstly  confirmed  in NGC~1068  (Miller  et  al.  1991), the  central
emission  in Seyfert~2  is scattered  by free  electrons in  a conical
structure placed along  the axis of the torus,  the so-call ionisation
cones, generating  a polar scattering  spectrum with a  position angle
perpendicular  to  the scattering  cone.  In  contrast, the  polarised
spectrum  of Seyfert~1  galaxies does  not conform  with  simple polar
scattering,  exhibiting a wide  diversity of  polarisation properties.
These results  state that  the geometry of  a single  polar scattering
region is  incomplete to explain  simultaneously all types  of Seyfert
polarised spectra.  Based on a study of 36 Seyfert~1 galaxies Smith et
al.   (2002) proposed  a model  in  which the  broad-line emission  is
originated in a rotating disk  and scattered in two different regions:
the  classical  {\it  ionisation   cones}  responsible  of  the  polar
scattering  and an  equatorial  scattering region  located within  the
torus  and  co-planar with  the  rotating  disc.   Therefore, the  two
extreme cases are  the face--on Seyfert~1 which exhibit  a null or weak
polarisation  as a  result  of cancellation  of  polar and  equatorial
scattering  and the Seyfert~2,  dominated by  polar scattering  as the
equatorial scattering region is completely obscured. In Seyfert~1 with
an  intermediate line  of  sight angle,  both  scattering regions  are
visible but in general equatorial polarisation dominates.

Interestingly,  a   peculiar  type  of   Seyfert~1  galaxies  exhibits
polarised spectra  similar to those  of Seyfert~2, i.e.   dominated by
polar scattering.   According to  the model proposed  by Smith  et al.
(2002),  these {\it  polar--scattered}  Seyfert 1  galaxies should  be
observed at  an inclination comparable  with the torus  opening angle,
and suffer therefore only a moderate extinction through the torus rim.
Smith et al.  (2004) estimate  that between 10\% and 30\% of Seyfert~1
galaxies are dominated by polar scattering.  Their spectropolarimetric
observations identified twelve of  this type of objects.  These twelve
objects constitute  the complete sample of  all known polar--scattered
Seyfert~1 galaxies.

X--ray studies  of these distinctive  {\it polar--scattered} Seyfert~1
galaxies are a powerful tool to  prove the basis of the model proposed
by Smith  et al.  (2004) and  therefore to further test  the scheme of
Unified Models for  AGNs. We have analysed the  eight objects included
in  the Smith  et al.  (2004) sample  with available  X-ray  data.  We
present for  the first  time, X-ray analysis  performed with  \xmm\ of
four them  (Fairall 51, Mrk  704, ESO 323-G077, and  IRAS 15091-2107).
The results  of the four remaining  ones (Mrk 231, NGC  3227, Mrk 766,
and NGC 4593) have been obtained from the literature. In the following
section, we  present the results on  the analysis of  the four objects
observed by  the first time by  \xmm.  In Section~\ref{sect:results},
we  combine   our  results  with  the  four   already  published  {\it
polar--scattered} Seyfert  1 galaxies  and address the  conclusions of
this work.

\section{Data Analysis}

We analysed for  the first time the \xmm\ observation  of four of the
objects  included in the  Smith et  al. (2004)  sample. All  data were
processed with  the standard {\it Science  Analysis System}, SASv7.0.0
and  using the  most updated  calibration files  available  in January
2007. The \epic\ event lists were  filtered to ignored periods of high
background  flaring following  the  method proposed  by Piconcelli  et
al. (2004).  According to the {\it SAS} task {\it epaplot}, no sign of
pile-up  was  detected for  none  of  the  objects. The  spectra  were
extracted from circular regions centred on the maximum emission of the
source. The  background regions  were extracted from  circular regions
located closed to the source and free of any contamination source.  In
all cases,  both {\it  MOS} spectra were  combined to obtain  a higher
signal--to--noise.  After  background subtraction, both  spectra, i.e.
the \pn\  and the {\it MOS1-2}  combined, were grouped  such that each
bin contains at  last 30 counts per bin. We are  therefore able to use
the modified  $\chi^2$ technique in  the spectral analysis.   When the
observations presented  enough signal--to--noise, {\it  RGS} data were
analysed.

We  have  performed the  spectral  analysis  of  the data  using  {\it
Xspec}~v.12.3.0. All  parameters are given in the
object rest  frame.  We  assumed a flat  $\Lambda CDM$  cosmology with
($\Omega_M,\Omega_\Lambda$)=(0.3,0.7)  and  a  value  for  the  Hubble
constant of 70 kms$^{-1}$ (Bennett  et al. 2003). The errors quoted in
this   paper    refer   to    the   90\%   confidence    level   (i.e.
$\Delta\chi^2$=2.71).  For the spectral  fitting of the \xmm\ data, we
have  considered the  absorption due  to the  Galaxy for  each object,
fixing the equivalent  Hydrogen column density to the  values given by
Dickey \& Lockman, (1990).

\begin{table*}
  \caption{Summary of the properties of the \xmm\ data analysis.}\label{tab:models}

\begin{minipage}{280mm}
 \begin{tabular}{@{}llllcc@{}}
  \hline
{\bf OBJECT}    &  \multicolumn{3}{c}{{\bf SOFT BAND}}  & \multicolumn{2}{c}{{\bf HARD BAND}} \\ 
                & \multicolumn{1}{c}{{\bf Cold Abs.$^1$}} & \multicolumn{2}{c}{{\bf Warm Abs.$^1$}} &  \multicolumn{1}{c}{{\bf Power Law}} &  \multicolumn{1}{c}{{\bf Broad Fe Line}} \\
                & N$_{H}^{c}$  & LogU & N$_{H}^{w}$               & $\Gamma$                                  \\\hline
%                &  $\times 10^{22}$ cm$^{-2}$ & & $\times 10^{22}$ cm$^{-2}$                                        \\\hline

Fairall 51      & 1.6$\pm$0.2 & -0.8 & 0.1 & 2.01$\pm0.12$ & $\checkmark$ \\
                &             & 0.12 & 7.9 & \\
                &             & 1.4  & 5.0 & \\
Mrk 704         & 2.5$\pm$1.5 & -0.3 & 0.4 & 1.66$\pm0.16$ & $\checkmark$ \\
                &             &  1.8 & 2.5 & \\
ESO 323-G77     & 6.0$\pm$0.3 &  -   & -   & 1.89$\pm0.05$ & $\checkmark$ \\
IRAS 15091-2107 & 0.7$\pm$0.2 & -0.03& 0.2 & 1.83$\pm0.06$ & $\times$ \\ \hline

\end{tabular}

\end{minipage}
\footnotesize{   Note:   (1)   Both   cold  (N$_{H}^{c}$)   and   warm
(N$_{H}^{c}$) Hydrogen equivalent column density are given in units of
10$^{22}$ cm$^{-2}$.}
\end{table*}

\subsection{The hard band spectrum}

\begin{figure}[!t]
  \includegraphics[width=60mm, angle=-90]{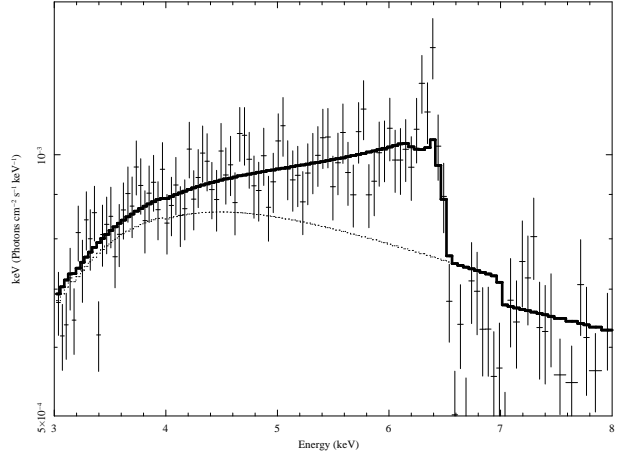}
  \caption{ESO~323-G077  \pn\  hard band  observed  spectrum and  best
  fitting  model. The data  show the  presence of  a broad  iron line
  compatible with being originated in a rotating accretion disc.}
  \label{fig:eso}
\end{figure}

We fitted the \epic\ X-ray spectra above 2~keV (but also excluding the
Fe line energy region) of the  four objects with a power law model. In
all cases, the emission model provides a satisfactory fit to the data.
The   values  of   the  index   of  the   power  law   are   given  in
Table~\ref{tab:models}.   All values  are in  good agreement  with the
typical results  of Type I  objects, e. i. $\Gamma=1.89\pm0.11$  for a
sample of 40 Palomar-Green QSO, (Piconcelli et al.  2005). In three of
the objects, we find hints of the presence of broad emission line.  In
the case  of ESO~323-G077, we have proved  the relativistic broadening
of the iron  emission line, probably originated in  the accretion disc
of  a Kerr  Black Hole  (Jim\'enez-Bail\'on  et al.   2007 in  prep.).
Fig.~\ref{fig:eso} shows  the ESO~323-G077 hard band  spectrum and the
fitted model evidencing the presence  of the broad Fe line. Therefore,
in the  hard band, the  objects studied present properties  typical of
Seyfert 1 galaxies.

\subsection{The soft band spectrum}

\begin{figure}[!t]
  \includegraphics[width=80mm]{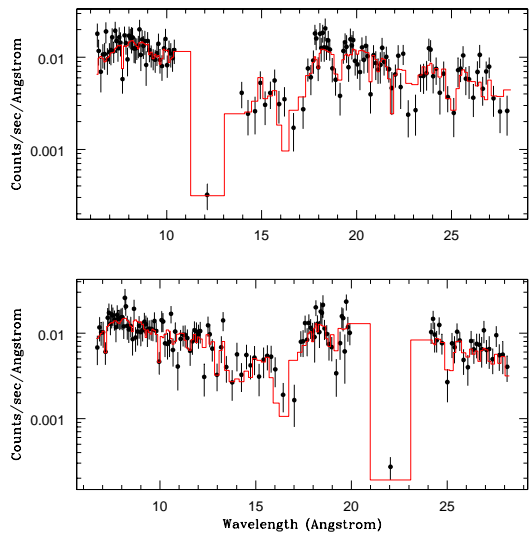}
  \caption{Fairall 51  {\it RGS}  observed  spectrum and  best
  fitting  model.}
  \label{fig:f51}
\end{figure}

We  also  investigated  the  emission of  the  {\it  polar--scattered}
Seyfert 1 galaxies in the soft band. In particular, we have found that
all  four galaxies present  evidence of  cold and/or  warm absorption.
Observed values of the  equivalent Hydrogen column densities are given
in Table~\ref{tab:models}.  The measured  N$_{H}^{c}$ are in all cases
of the order of  $10^{22}$~cm$^{-2}$. This result diverges from common
properties  of the  optically bright  Type  I objects,  for which  the
incidence  of  cold  absorption  not  explained by  the  host  galaxy,
i.e.  $>10^{21}$~cm$^{-2}$  is  $\le$5\%  (Piconcelli  et  al.   2005).
Bearing in mind the Smith et  al. (2004) model, we tested the possible
presence of warm  absorption in the objects. The
imprints  of warm absorbers  in the  soft X-ray  band are  observed in
about  half of  the Type  I AGN  (Piconcelli et  al. 2005,  Blustin et
al. 2007).   Using the {\it Phase}  model (Krongold et  al.  2003), we
found  evidence of  warm absorption  in three  of the  four  Seyfert 1
analysed.   The most  outstanding case  is found  for  Fairall~51 (see
Figure~\ref{fig:f51}) in which three different warm absorbers could be
detected  and parametrised.   The properties  of all  detected ionised
absorbers are given in Table~\ref{tab:models}.

\section{Results}\label{sect:results}

The results  on the analysis  performed on the four  observed galaxies
can be summarised as follows:

\begin{itemize}

\item The properties of the  galaxies in the hard X-ray band indicate
that we are directly observing the central AGN. The index of the power
law and measured luminosities are  typical of Type I AGN. This result
is  also supported by  the presence  of relativistically  broaden iron
lines in three out of the four observed objects. 

\item Conversely, in the soft energy band, the X-ray properties of the
studied objects  resemble to those  of absorbed AGN.   Cold absorption
incompatible with being originated in the host galaxy was reported for
all  the objects.  Evidence of  warm absorption  was measured  in also
three of the Seyfert 1 galaxies in the sample.

\end{itemize}

Similar general properties are observed for the {\it polar--scattered}
Seyfert  1 galaxies with  already published  results.  Details  on the
X-ray data  analysis of the objects can  be found in for  Mrk 231, for
NGC~3227, for Mrk 766, and for NGC 4593.

In   summary,  the   analysis  of   new  \xmm\   data  of   four  {\it
polar--scattered} Seyfert  1 galaxies combined  with published results
of other four  objects support the Smith et  al. (2004) interpretation
for these peculiar type of Seyfert galaxies.

\end{document}